\documentstyle[12pt]{article}
\makeatletter

\ifcase \@ptsize
\or 
\or 
\fi
\advance\textheight by \topskip

\ifcase \@ptsize
\or 
\or 
\fi

\renewcommand{\theequation}{\arabic{equation}}
\def\section{\@startsection {section}{1}{\z@}{-3.5ex plus -1ex minus
 -.2ex}{2.3ex plus .2ex}{\large\bf\centering}}
\def\subsection{\@startsection{subsection}{2}{\z@}{-3.25ex plus -1ex minus
 -.2ex}{1.5ex plus .2ex}{\sc}}
\def\@cite#1#2{\nolinebreak$^{[\scriptstyle #1\if@tempswa , #2\fi]}$}
\def\@citex[#1]#2{\if@filesw\immediate\write\@auxout{\string\citation{#2}}\fi
  \def\@citea{}\@cite{\@for\@citeb:=#2\do
    {\@citea\def\@citea{,\penalty\@m}\@ifundefined
       {b@\@citeb}{{\bf ?}\@warning
       {Citation `\@citeb' on page \thepage \space undefined}}%
{\csname b@\@citeb\endcsname}}}{#1}}
\@definecounter{footnote}

\gdef\@publabel{\hfil}
\gdef\@pubdate{\null}
\gdef\@pubnumber{\null}
\gdef\@author{\null}
\gdef\@title{\null}
\gdef\@abstract{\null}
\long\def\pubdate#1{\gdef\@pubdate{#1}}
\long\def\pubnumber#1{\gdef\@pubnumber{#1}}
\long\def\publabel#1{\gdef\@publabel{#1}}
\long\def\author#1{\gdef\@author{#1}}
\long\def\title#1{\gdef\@title{#1}}
\long\def\abstract#1{\gdef\@abstract{#1}}
\def\titlerelax{
\let\maketitle\relax
\let\settitleparameters\relax
\let\consolidatetitle\relax
\let\inittitlepage\relax
\let\finishtitlepage\relax
\let\titlepagecontents\relax
\let\multithanks\relax
\let\titlebaselines\relax
\let\@makepub\relax
\let\@maketitle\relax
\let\@makeauthor\relax
\let\@makeabstract\relax
\let\@maketitlenote\relax
\let\thanks\relax
\let\titlerelax\relax}
\def\titleclean
{\gdef\@titlenote{}
\gdef\@abstract{}
\gdef\@author{}
\gdef\@title{}
\gdef\@pubdate{}\gdef\@pubnumber{}\gdef\@publabel{}
\gdef\@dpublabel{}
}

\def\@makepub{\vbox to \z@{\hbox to \textwidth{\hfill
\@publabel \hfill
\llap{\parbox[t]{0.25\textwidth}{\raggedleft\@pubnumber}}}%
\vss}}
\def\@maketitle{\vskip 60pt \begin{center}
 {\LARGE \@title \par}
 \end{center}}
\def\@makeauthor{{\def\and{\smallskip {\normalsize \rm and\smallskip}}
\long\def\address##1{{\def\and{\\and\\}\medskip
                {\small \it \\##1\\}
}}
{\centering
 \vskip 1.5em
 \large \lineskip .75em
 \@author}
 \par}}
\def\@makedate{\vskip 1.5em
 {\raggedright \small \noindent\@pubdate \par}}
\def\@makeabstract{\vskip 1.5em
{\small
\begin{center}
{\bf ABSTRACT\vspace{-.5em}\vspace{0pt}}
\end{center}
\quotation \@abstract \endquotation}
\vspace{-1em}}
\def\maketitle{
\let\footnotesize\small \setcounter{page}{1}
\@makepub
\@maketitle
\@makeauthor
\@makeabstract
\@thanks
\@makedate
\setcounter{footnote}{0}
}

\begin{document}
\newcommand{\hsp}{\hspace{0.08in}}
\newcommand{\eqn}{\begin{equation}}
\newcommand{\e}{\end{equation}}
\bibliographystyle{npb}

\pubnumber{DAMTP--95--2\\ hep-th/9501079\\}

\title{Lattice quantization of Yangian charges}
\author{N. J. MacKay\footnote{Supported by a PPARC
fellowship}\footnote{n.j.mackay@damtp.cambridge.ac.uk}
\address{Dept of Applied Maths and Theoretical Physics, \linebreak
         Cambridge University, \linebreak
         Cambridge, CB3 9EW, UK}}
\abstract{
By placing theories with Yangian charges on the lattice in the
analogue of the St Petersburg school's approach to the sine-Gordon
system, we exhibit the Yangian structure of the auxiliary algebra, and
explain how the two Yangians are related.
}

\maketitle
\baselineskip 18pt
\parskip 15pt
\parindent 10pt

\section{Introduction}

In 1990 Bernard\cite{Bern}, building on work by
L\"uscher\cite{L}, showed that the algebra of non-local charges
in 1+1-D quantum field theories with curvature-free conserved currents
is precisely Drinfeld's Yangian algebra\cite{D}. The naive
definition of the first non-local charge is divergent, and must
be regularized; Bernard did this on the continuum. The coproduct
gives the action of the charges on asymptotic states, and
conservation of the charges allows one, in principle, to determine
the scattering matrix. The classical $\hbar\rightarrow 0$
limit\cite{cyang} of the charge algebra retains the Yangian
structure in both the Poisson brackets and (a classical definition
of) the coproduct (see appendix).
No spectral parameter or auxiliary algebra appears
in this procedure, and its use was therefore presented as an
alternative to the quantum inverse scattering method.

Recall now how Drinfeld introduced the
Yangian, as the unique quantum deformation of a Lie algebra consistent
with a Casimir-like classical $r$-matrix.
This Yangian is the quantum version of the auxiliary
algebraic structure introduced when integrability of a classical
equation (in this case, conservation and curvature-freedom of the
current) is expressed through a Lax pair, which involves a spectral
parameter.

Drinfeld's mathematical intuition may be contrasted
with the physical intuition whereby Kulish and Reshetikhin\cite{KulR}
discovered the quantum deformation of $su(2)$. The quantum
inverse scattering method (QISM) of the St Petersburg school takes the
classical Lax system and its monodromy matrix, which diverges upon
quantization, and regularizes the quantum system by placing it
on a spatial lattice, defining the spatial lattice Lax operator as the
quantization of the monodromy matrix over one lattice step, which
is finite. Making the quantum analogue of the classical $r$-matrix
relation soluble, however, may require that the (`auxiliary') algebraic
structure of the classical Lax pair be deformed. For sine-Gordon
theory the classical $su(2)$ algebra becomes
what is now known as $su_q(2)$.

It is therefore interesting to see whether we can
relate the Yangian charges, which act on the `quantum' space
of QISM terminology, to the Yangian of the `auxiliary' space,
by quantizing on the lattice the Lax system of models with
Yangian charges. We find that what is apparently a loop
algebra acting on the auxiliary space must indeed be replaced by
a Yangian, and in doing so place the Yangian on the same footing
as $su_q(2)$ in terms of the intuitions and historical development
of QISM.

\section{Auxiliary Yangian structure}

Consider a classical current which is conserved,
$$
\partial^{\mu}j_{\mu}(x,t)=0 \hsp,
$$
curvature-free,
$$
\partial_{\mu} j_{\nu} - \partial_{\nu} j_{\mu} + [j_{\mu},j_{\nu}] =
0 \hsp, $$ and Lie algebra valued, $$ j_{\mu}(x,t) =
t^{a}\,j_{\mu}^{a}(x,t) $$ (where the $t^{a}$ generate a Lie
algebra ${\cal A}$, $[t^a,t^b]=f^{abc}t^c$). These conditions are
equivalent to the vanishing of the curvature $$ [\partial_0 + L_0
\,,\, \partial_1 + L_1 ] = 0 $$ of the Lax pair $$ L_\mu(x,t;u) =
{1 \over 1-u^2} \left( j_\mu(x,t) + u \epsilon_\mu^{\;\nu}
j_\nu(x,t) \right) $$ (with convention $\eta_{00}=-\eta_{11}=1$)
and thus to the solubility of
$(\partial_1+L_1)T=0=(\partial_0+L_0)T$. The equation in the
spatial derivative, \eqn\label{lax} \left(
\partial_x + L_1(x;u) \right) T(x,y;u) = 0 \hsp,\
\e
has the formal solution \eqn\label{tm} T(x,y;u) = {\bf P} \exp
\left( -\int_y^x L_1(\xi;u) \,{d\xi} \right) \hsp,
\e
where {\bf P} denotes path ordering on a path at a fixed time, and $t$
is a suppressed label. The integrability of the system is expressed by
\eqn\label{cr}
\hspace{0.3in}\left\{ T(u) \stackrel{\otimes}{,} T(u')
 \right\} = \left[
r(u,u'), T(u) \otimes T(u')
\right]\;,
\e
where
$$
T(u)=T(\infty,-\infty;u) \;,\hspace{0.2in} r(u,u') = {1\over u'-u}
t^a\otimes t^a \;,
$$
and the tensor product is in the auxiliary space; taking the trace
of this relation gives an infinity of charges in involution.
The passage to the quantum theory is beset by problems due to
divergences in (\ref{lax}) and (\ref{tm}) which arise from products
of quantum operators valued at the same point. The solution of the
St Petersburg school\cite{FST} is to put the theory on a lattice
and define
the quantum theory by its lattice transfer matrix, which is fixed
by the requirements of integrability, {\em i.e.\ }that
\eqn\label{R}
R(u,u') \, L_N^1(u) \,  L_N^2(u')  \, = \,
L_N^2(u') \, L_N^1(u) \, R(u,u')
\e
be soluble, where $L_N$ is the one-step lattice transfer matrix and
$$
L_N^1(u)=L_N(u) \otimes 1 \hspace{0.5in}
L_N^2(u')=1 \otimes L_N(u') \hsp,
$$
and of
\eqn\label{R2}
R(u,u') \, T^1(u) \,  T^2(u')  \, = \,
T^2(u') \, T^1(u) \, R(u,u')
\e
having the correct classical limit (\ref{cr})
(where $T$ is the full lattice transfer
matrix, $\prod_{N=-\infty}^\infty L_N$).

The natural ansatz\cite{KS2,qtlo} for $L_N$ is to set $$ L_N
\equiv T((N-1/2)\Delta,(N+1/2)\Delta;u) = {\bf P} \exp \left(
-\int_{(N-{1\over 2})\Delta}^{(N+{1\over 2})\Delta}
 L_1(\xi;u) \,{d\xi} \right)
$$ where the lattice points are $x_N=N\Delta$. We can then
calculate $L_N$ using the canonical quantization of $L_1(x;u)$,
and find, writing the operator products as the sum of a regular
part and a commutator, that it is finite. Using light-cone
components of the current, $j_\pm\equiv{1\over2}(j_0 \pm j_1)$, we
take the commutation relations of the lattice currents $$ j_{N\pm}
\equiv \int_{(n-{1\over 2})\Delta}^{(n+{1\over 2})\Delta}
j_\pm(\xi)\,{d\xi} $$ to be $$ [\,j_{N\pm}\,,\,j_{M\pm}\,]\; = \;
-i\hbar \,g\, \delta_{NM} j_{N\pm} \, - \, {ik \Delta \over
\lambda} \delta_{NM} C_2 \hsp, $$ where $k$ is an unknown constant
with the dimensions of energy, $\lambda$ a coupling constant,
$C_2=t^at^a$ and $\delta^{ad}g=f^{abc}f^{dcb}$. (The classical
limit has to be very carefully taken, with the space component of
the current rescaled by a factor of $\hbar/\Delta$ in the way
suggested by Faddeev and Reshetikhin\cite{FR}, in order to get the
Poisson brackets of the principal chiral model ${\cal L} =
\lambda^{-1} \int {\rm Tr}(
\partial g^{-1} \partial g)$.) For the moment we shall set $k=0$, and describe what
happens for non-zero $k$ later.

We then obtain $$ L_N(u) = 1 \,+\, \sum_{n=0}^\infty \left(\,
\left({1\over 1-u}\right)^{n+1} j_{N-}^a \,+\, \left({-1\over
1+u}\right)^{n+1} j_{N+}^a \,\right) (-1)^n t^a_n + {\cal
O}(\epsilon^2) $$ where $$ t^a_n=(\epsilon g)^n t_a \; ,
\hspace{0.15in} \epsilon={i\hbar\over 2}\;. $$ (Note that our
approach has been to treat $\epsilon g$ as a loop variable,
independent of $\epsilon$; none of its powers have been absorbed
into the ${\cal O}(\epsilon^2)$ terms.) We shall see in the next
section that the currents, when summed in the monodromy matrix,
give charges which form a Yangian (acting on the {\em quantum}
space). Similarly the $t_n^a$, we shall find when we work out
$L_N^1L_N^2$, must be deformed from loop algebra generators into
the generators of another Yangian (acting on the {\em auxiliary}
space), and it follows (upon expanding and re-summing the $(1\pm
u)^{-n-1}$ terms) that $$ L_N(u) = 1 \,-\, T_u \sum_{n=0}^\infty
 t_n^a j_{N\mu}^a \;+\; {\cal O}(\epsilon^2) \;, $$
where $\mu=0$ for $n$ odd and $\mu=1$ for $n$ even, and $T_u$ acts
on $t_n^a$ according to (\ref{AMh}). Working out $L_N^1L_N^2$ in
the same way we find $$ L_N^1(u)L_N^2(u') = 1\otimes 1 -T_u\otimes
T_{u'} \sum_{n=0}^\infty \Delta(t^a_n) j_{N\mu}^a + {\cal
O}(\epsilon^2) $$ (with $\Delta(t^a_n)$ given by (\ref{deltah})),
a remarkably neat result, and one of the main calculations of this
paper. The requirement that (\ref{R}) be soluble is satisfied if
(\ref{cocomm}) holds, and thus if the $t_n^a$ form a Yangian. We
believe the reverse implication to be true also, but it does not
automatically follow, and cannot be proved without knowing the
${\cal O}(\epsilon^2)$ terms (which will depend on the
quantization prescription) exactly.

Now let us examine what happens for non-zero $k$. After some
calculation we find that $L(u) \rightarrow L(u - {c\over 1-u})$,
where $c=\epsilon k \Delta C_2 / \lambda\hbar$, and thence $$
L_N^1(u) L_N^2(u') = 1\otimes 1 + T_{u-{c\over 1-u}}\otimes
T_{u'-{c\over 1-u'}} \sum_{n=0}^\infty \Delta(t^a_n) j_{N\mu}^a +
{\cal O}(\epsilon^2)\hsp, $$ up to terms of order $\epsilon$
proportional to $t^a\otimes t^a$, which commute with a
group-invariant $R$-matrix.

\section{Yangian charges}

The conserved charges may now be extracted as coefficients of
powers of $u$ in $T= \prod_{N=-\infty}^\infty L_N$, and we find
that $$ T(u)= \exp\left( - \sum_{n=0}^\infty {1\over u^{n+1}}
\sum_{p=0}^n \left({n \atop p} \right)(-1)^p Q_p^a t_{n-p}^a
\right) \;+\;{\cal O}(\epsilon^2) \hsp, $$ where
\begin{eqnarray*}
Q_0^a  & = & \sum_{N=-\infty}^\infty j_{N0}^a \\ Q_1^a & = &
\sum_{N=-\infty}^\infty \left(\;j_{N1}^a \,+\, {1\over 2}f^{abc}
 j_{N0}^b \sum_{N'=-\infty}^{N-1} j_{N'0}^c \;\right)
\end{eqnarray*}
and higher charges are defined by iteration, equivalent
to the (`algebra-valued'\cite{deVEM1}) charges produced by
the usual classical procedure\cite{BIZZ}.
It follows that the charges satisfy (\ref{mult},\ref{YSerre}), whilst the
coproduct can be defined as for Bernard's continuum
charges or (equivalently) by examining the classical
limit\cite{cyang,LP}, and is indeed (\ref{delta}).

Now recall\cite{Bern,L} that a Lorentz boost $L_\theta$ which adds
$\theta$ to the rapidity of a state on which the charges are
measured has the effect $$ L_\theta:\;Q_0^a \mapsto Q_0^a
\;,\hspace{0.2in} Q_1^a \mapsto Q_1^a - {\hbar g\over 4i\pi}
\theta Q_0^a = T_{{\hbar g\over 4i\pi}\theta}  \, Q_1^a \;. $$
(This is easily seen to apply in the same way to the lattice
charges.) It may be checked that \eqn\label{TL} L_\theta T(u) =
T\left( u - {\hbar g\over 4i\pi}\theta\right) \hsp.
\e

\section{Overview}

None of these results is surprising if one takes the
view\cite{deVEM1}
that the most efficient way to construct an integrable model
in this way is to take an abstract, formal $R$-matrix and
represent it on combinations of auxiliary and quantum spaces
as required. The monodromy matrix is then just the formal $R$-matrix
evaluated on one quantum and one auxiliary space, and it
is necessarily true in this approach that $T$ depends additively
on $u$ and $\theta$, as in (\ref{TL}).
The Yang-Baxter equation (YBE) on three
quantum spaces gives factorization of the scattering matrix;
on two quantum and one auxiliary space, conservation of conserved
charges (this is well described in the literature\cite{Bern,cn});
on two auxiliary and one quantum space, the integrability condition
(\ref{R2}); and on three auxiliary spaces, the YBE for the $R$-matrix.
The underlying algebra, the Yangian, is the same whether acting
on the quantum or the auxiliary space, and the four possibilities
just outlined are (\ref{cocomm}) and (\ref{YBE}) for its two manifestations.
That all the results necessary for this approach can be proved
in the explicit QISM-motivated construction above is rather
satisfying, and should perhaps be taken as an indication that
any consistent lattice quantization of an integrable 1+1-D
field theory must be performed in this way.

\newpage
\appendix{\centerline{\large \bf Appendix: The Yangian algebra}}
\renewcommand{\theequation}{Y\arabic{equation}}

\setcounter{equation}{0}

The Yangian\cite{D} ${\rm Y}({\cal A})$ of a Lie algebra ${\cal
A}$
is generated by
$I_0^a,I_1^a$ satisfying
\begin{eqnarray}\nonumber
[I_0^a,I_0^b] & = & \alpha f^{abc} I_0^c \\
\label{mult} [I_0^a,I_1^b] & = & \alpha f^{abc} I_1^c
\end{eqnarray}
where $f^{abc}$ are the structure constants of ${\cal A}$.
The bi-algebra structure is given by the trivial co-unit
$$
\epsilon(I_0^a)=0=\epsilon(I_1^a) \hsp, \hspace{0.25in} \epsilon(1)=1
$$
and the coproduct
\begin{eqnarray}\nonumber
\Delta(I_0^a) & = & 1\otimes I_0^a + I_0^a \otimes 1 \\
\label{delta}
\Delta(I_1^a) & = & 1\otimes I_1^a + I_1^a \otimes 1 + {\alpha'\over
2}f^{abc}
I_0^c \otimes I_0^b \hsp.
\end{eqnarray}
Requiring that (\ref{delta}) be a homomorphism gives
\eqn\label{YSerre}
{1\over2} f^{d[ab} [I_1^{c]},I_1^d] =
 \alpha\alpha'^2 a_{abcpqr} I_0^{(p}I_0^{q}I_0^{r)}
\hsp,
\e
where $[\,]$ and $(\,)$ denote (anti-)symmetrization on the enclosed
indices and
$$
a_{abcpqr} = {1\over{24}} f^{ap i} f^{bq j}f^{cr k}f^{ijk} \hsp.
$$
(This applies for ${\cal A} \neq sl(2)$; for ${\cal A} = sl(2)$
see Drinfeld.)
One should think of ${\rm Y}({\cal A})$ as being generated by a
series of generators in adjoint representations of ${\cal A}$ at grades
$0,1,2,..$, with the $I_0^a$ and $I_1^a$ being simply the first two
sets, at grades
$0$ and $1$ respectively (and then $\alpha'$ formally has grade 1).
The condition (\ref{YSerre}) then
gives a constraint on the construction of higher grade generators from
products of $I_1$s. Specifically, defining
\eqn\label{higher}
I_p^a \equiv {1\over g\alpha} f^{abc} [I_1^c, I_{p-1}^b] \hsp,
\e
we find, applying (\ref{delta}) to (\ref{higher}), that
\eqn\label{deltah}
\Delta(I_p^a) = I_p^a \otimes 1 + 1 \otimes I_p^a
+ {\alpha'\over 2} f^{abc}\left( I_{p-1}^c \otimes I_0^b +
I_{p-2}^c \otimes I_1^b +\;\dots\;+
I_{1}^c \otimes I_{p-2}^b+I_{0}^c \otimes I_{p-1}^b \right) +
{\cal O}(\alpha'^2) \hsp.
\e
We have given ${\rm Y}({\cal A})$ in terms of two deformation
parameters $\alpha,\alpha'$, one of which may be scaled out.
In the charge algebra we have $I_p^a\equiv Q_p^a$ and
$\alpha=i\hbar,\alpha'=1$, whilst
in the auxiliary Yangian we have $I_p^a\equiv t_p^a$ and
$\alpha=1,\alpha'=i\hbar$.

To make ${\rm Y}({\cal A})$ a Hopf algebra we also need the
antipode map $$ s(I_0^{a})=-I_0^{a}\hspace{0.3in}{\rm
and}\hspace{0.3in} s(I_1^{a})=-I_1^{a}+{1\over 2}f^{abc} I_0^{b}
I_0^{c} \hsp, $$ which is an algebra antihomomorphism. This
satisfies $$ \cdot(s\otimes 1) \sigma\circ\Delta = \cdot(1\otimes
s)\sigma\circ\Delta = 0 $$ where $\cdot$ denotes multiplication
(in the enveloping algebra) and $\sigma:x\otimes y \mapsto
y\otimes x$ is the transposition operator. (This differs from the
usual axiom in that we have used $\sigma\circ\Delta$ rather than
$\Delta$; this has been done to maintain the conventional ordering
of the currents in the quantum definition of the non-local
charge.)

Further, the Yangian has an automorphism $T_{\lambda}$,
$\lambda\in {\cal C}$, given by $$ T_{\lambda}:I_1^{a} \mapsto
I_1^{a} - \lambda I_0^{a} \hspace{0.2in} {\rm and}
 \hspace{0.2in} T_{\lambda}:I_0^{a} \mapsto I_0^{a} \hsp,
$$ and more generally \eqn\label{AMh} T_{\lambda}:I_p^a \mapsto
\sum_{r=0}^p  \left({p \atop r}\right) (-\lambda)^r I_{p-r}^a
\hsp.
\e
There then exists a formal $R$-matrix, ${\cal R}(\lambda)$, with
$(T_{\lambda_1} \otimes T_{\lambda_2} ) {\cal R}(\lambda)={\cal
R}(\lambda+\lambda_1-\lambda_2)$, satisfying \eqn\label{cocomm}
(1\otimes T_{\lambda})\sigma\circ\Delta(x) = {\cal
R}(\lambda)^{-1}(1\otimes T_{\lambda})\Delta(x){\cal R}(\lambda)
\hspace{0.3in}(x\in Y({\cal A})) \hsp;
\end{equation}
this ${\cal R}$ satisfies the Yang-Baxter equation,
\eqn\label{YBE}
{\cal R}_{12}(\lambda){\cal R}_{13}(\lambda+\lambda')
{\cal R}_{23}(\lambda')
=
{\cal R}_{23}(\lambda'){\cal R}_{13}(\lambda+\lambda'){\cal
R}_{12}(\lambda) \;,
\e
and $$ {\cal R}_{21}(\lambda){\cal R}_{12}(-\lambda)=1 \hsp. $$ In
the auxiliary algebra $\lambda=u$, whilst in the charge algebra
$\lambda= {\hbar g\over 4i\pi} \theta$, this scaling being fixed
by the definition of the charges or (in the exact $S$-matrix
approach) by the requirement of crossing symmetry.

\end{document}